\documentclass[english,aps,superscriptaddress]{revtex4}
\usepackage[T1]{fontenc}
\usepackage[latin9]{inputenc}
\usepackage{mathrsfs}
\usepackage{amsbsy}
\usepackage{esint}

\makeatletter
\@ifundefined{textcolor}{}
{%
 \definecolor{BLACK}{gray}{0}
 \definecolor{WHITE}{gray}{1}
 \definecolor{RED}{rgb}{1,0,0}
 \definecolor{GREEN}{rgb}{0,1,0}
 \definecolor{BLUE}{rgb}{0,0,1}
 \definecolor{CYAN}{cmyk}{1,0,0,0}
 \definecolor{MAGENTA}{cmyk}{0,1,0,0}
 \definecolor{YELLOW}{cmyk}{0,0,1,0}
}

\makeatother

\usepackage{babel}
\begin{document}

\title{Magnetic moment, vorticity-spin coupling and parity-odd conductivity
of chiral fermions in 4-dimensional Wigner functions}

\author{Jian-hua Gao}

\affiliation{Shandong Provincial Key Laboratory of Optical Astronomy and Solar-Terrestrial
Environment, Institute of Space Sciences, Shandong University, Weihai,
Shandong 264209, China}

\author{Qun Wang}

\affiliation{Interdisciplinary Center for Theoretical Study and Department of
Modern Physics, University of Science and Technology of China, Hefei,
Anhui 230026, China}

\affiliation{Physics Department, Brookhaven National Laboratory, Upton, New York
11973-5000, USA}
\begin{abstract}
We demonstrate the emergence of the magnetic moment and spin-vorticity
coupling of chiral fermions in 4-dimensional Wigner functions. In
linear response theory with space-time varying electromagnetic fields,
the parity-odd part of the electric conductivity can also be derived
which reproduces results of the one-loop and the hard-thermal or hard-dense
loop. All these properties show that the 4-dimensional Wigner functions
capture comprehensive aspects of physics for chiral fermions in electromagnetic
fields.
\end{abstract}
\maketitle
\textit{1. Introduction}. Significant progresses have been made in
understanding the dynamics of chiral fermions in electromagnetic fields.
This is particularly interesting in high energy heavy ion collisions
where very strong magnetic fields can be created. The magnetic fields
are so strong that quarks can be polarized and their momentum directions
are parallel or anti-parallel to the magnetic field depending on quark
chiralities and charges. Quarks with the same charge tend to move
in the same direction. Any imbalance in the number of right-handed
and left-handed quarks as a consequence of topological configurations
of gauge fields may lead to such a charge separation effect which
can be tested in experiments \cite{Abelev:2009ac}. This is termed
as the Chiral Magnetic Effect (CME) \cite{Kharzeev:2007jp,Fukushima:2008xe}.
The Chiral Vortical Effect (CVE) is also an accompanying effect due
to rotation in a relativistic and charged fluid \cite{Son:2009tf,Kharzeev:2010gr}.
The interplay of chiral magnetic and chiral separation effects may
lead to a phenomenon called the Chiral Magnetic Wave \cite{Burnier:2011bf},
whose vortical counter part is the Chiral Vortical Wave \cite{Jiang:2015cva}. 

Kinetic theory is an important tool to describe these phenomena in
phase space of chiral fermions. The Abelian Berry potential takes
an important role in 3-dimensions (3D) kinetic approach in accommodation
of axial anomaly \cite{Son:2012wh,Stephanov:2012ki,Chen:2013iga}.
It has been shown that the CME, CVE and Covariant Chiral Kinetic Equation
(CCKE) can be derived in quantum kinetic theory from the Wigner function
in 4-dimensions (4D) in external electromagnetic fields \cite{Gao:2012ix,Chen:2012ca}.
The 3D chiral kinetic equation \cite{Son:2012wh,Stephanov:2012ki,Chen:2013iga}
can be obtained from the CCKE by integrating out the zero component
of the 4-momentum. 

In the 3D chiral kinetic equation, it has been shown that the fermion
energy is shifted by the interaction energy of magnetic moment with
the magentic field \cite{Son:2012zy}. The magnetic moment and spin
of fermions have relativistic origin \cite{Chen:2014cla,Duval:2014ppa,Manuel:2014dza}.
It is a natural conjecture that the magnetic moment should also emerge
in the covariant quantum kinetic approach in 4D Wigner functions.
In this paper, we will demonstrate the emergence of the magnetic moment
as well as spin-vorticity coupling in the framework of covariant quantum
kinetic theory based on 4D Wigner functions. We will also show that
the parity-odd part of electric conductivity (chiral magnetic conductivity)
can also be derived from 4D Wigner functions in linear response theory
with space-time varying electromagnetic fields. The result reproduces
the chiral magnetic conductivity of one loop \cite{Kharzeev:2009pj}
and hard-thermal or hard-dense loop (HTL or HDL) \cite{Braaten:1989mz,Laine:2005bt}
under proper approximations \cite{Son:2012zy,Manuel:2013zaa}. 

We adopt the same sign conventions for the fermion charge $Q$ and
the axial vector component of the Wigner function as in Refs. \cite{Vasak:1987um,Gao:2012ix,Chen:2012ca}. 

\textit{2. Wigner functions for chiral fermions}. In a background
electromagnetic field, the quantum mechanical anologue of a classical
phase-space distribution for fermions is the gauge invariant Wigner
function $W_{\alpha\beta}(x,p)$ which satisfies the equation of motion
\cite{Elze:1986qd,Vasak:1987um}, $\left(\gamma_{\mu}K^{\mu}-m\right)W(x,p)=0$,
where $x=(x_{0},\mathbf{x})$ and $p=(p_{0},\mathbf{p})$ are space-time
and energy-momentum 4-vectors. For the constant field strength $F_{\mu\nu}$,
the operator $K^{\mu}$ is defined by $K^{\mu}=p^{\mu}+i\frac{1}{2}\nabla^{\mu}$
with $\nabla^{\mu}=\partial_{x}^{\mu}-QF^{\mu\nu}\partial_{\nu}^{p}$.
The Wigner function can be decomposed in 16 independent generators
of Clifford algebra, whose coefficients $\mathscr{F}$, $\mathscr{P}$,
$\mathscr{V}_{\mu}$, $\mathscr{A}_{\mu}$ and $\mathscr{S}_{\mu\nu}$
are the scalar, pseudo-scalar, vector, axial-vector and tensor components
of the Wigner function respectively. For massless or chiral fermions,
the equations for $\mathscr{V}_{\mu}$ and $\mathscr{A}_{\mu}$ are
decoupled from other components of the Wigner function. 

\textit{3. Formal solution to quantum kinetic equations}. For chiral
(massless) fermions, we can rewrite the equations for $\mathscr{V}_{\mu}$
and $\mathscr{A}_{\mu}$ into those for right-handed $(s=+)$ and
left-handed ($s=-$) vectors $\mathscr{J}_{\mu}^{s}(x,p)$, 
\begin{eqnarray}
p^{\mu}\mathscr{J}_{\mu}^{s}(x,p) & = & 0,\nonumber \\
\nabla^{\mu}\mathscr{J}_{\mu}^{s}(x,p) & = & 0,\nonumber \\
2s(p^{\lambda}\mathscr{J}_{s}^{\rho}-p^{\rho}\mathscr{J}_{s}^{\lambda}) & = & -\epsilon^{\mu\nu\lambda\rho}\nabla_{\mu}\mathscr{J}_{\nu}^{s},\label{eq:wig-eq}
\end{eqnarray}
where $\mathscr{J}_{\mu}^{s}(x,p)$ are given by 
\begin{eqnarray}
\mathscr{J}_{\mu}^{s}(x,p) & = & \frac{1}{2}[\mathscr{V}_{\mu}(x,p)+s\mathscr{A}_{\mu}(x,p)].
\end{eqnarray}
One can derive a formal solution of $\mathscr{J}_{\mu}^{s}$ in Eq.
(\ref{eq:wig-eq}) by a perturbation method in powers of $\nabla^{\mu}$
and $F_{\mu\nu}$. The solutions at the zeroth and first order were
given by Ref. \cite{Gao:2012ix,Chen:2012ca} and can be cast into
the following form, 
\begin{eqnarray}
\mathscr{J}_{(0)s}^{\rho}(x,p) & = & p^{\rho}f_{s}\delta(p^{2}),\nonumber \\
\mathscr{J}_{(1)s}^{\rho}(x,p) & = & -\frac{s}{2}\tilde{\Omega}^{\rho\beta}p_{\beta}\frac{df_{s}}{dp_{0}}\delta(p^{2})-\frac{s}{p^{2}}Q\tilde{F}^{\rho\lambda}p_{\lambda}f_{s}\delta(p^{2}),\label{eq:1st-solution}
\end{eqnarray}
where $p_{0}\equiv u\cdot p$ with $u^{\mu}$ being the fluid velocity,
and $f_{s}$ are distribution functions of chiral fermions defined
by 
\begin{eqnarray}
f_{s}(x,p) & = & \frac{2}{(2\pi)^{3}}\left[\Theta(p_{0})f_{F}(p_{0}-\mu_{s})+\Theta(-p_{0})f_{F}(-p_{0}+\mu_{s})\right].\label{eq:dist}
\end{eqnarray}
Here $f_{F}(y)\equiv1/[\exp(\beta y)+1]$ is the Fermi-Dirac distribution
function, and $\beta=1/T$ and $\mu_{s}$ are the temperature inverse
and the chemical potential of chiral fermions with chirality $s$
respectively. We can decompose $\mu_{s}$ into the scalar and pseudo-scalar
parts, $\mu_{s}=\mu+s\mu_{5}$. We have used following formula, $\tilde{F}^{\rho\lambda}=\frac{1}{2}\epsilon^{\rho\lambda\mu\nu}F_{\mu\nu}$,
$\tilde{\Omega}^{\xi\eta}=\frac{1}{2}\epsilon^{\xi\eta\nu\sigma}\Omega_{\nu\sigma}$
and $\Omega_{\nu\sigma}=\frac{1}{2}(\partial_{\nu}u_{\sigma}-\partial_{\sigma}u_{\nu})$,
where $\epsilon^{\mu\nu\sigma\beta}$ and $\epsilon_{\mu\nu\sigma\beta}$
are anti-symmetric tensors with $\epsilon^{\mu\nu\sigma\beta}=1(-1)$
and $\epsilon_{\mu\nu\sigma\beta}=-1(1)$ for even (odd) permutations
of indices 0123, so we have $\epsilon^{0123}=-\epsilon_{0123}=1$.
Instead of $\Omega_{\nu\sigma}$, $\tilde{\Omega}^{\xi\eta}$, $F_{\mu\nu}$
and $\tilde{F}^{\rho\lambda}$, we will also use the vorticity vector
$\omega^{\rho}=\frac{1}{2}\epsilon^{\rho\sigma\alpha\beta}u_{\sigma}\partial_{\alpha}u_{\beta}$,
the electric field $E^{\mu}=F^{\mu\nu}u_{\nu}$, and the magnetic
field $B^{\mu}=\frac{1}{2}\epsilon^{\mu\nu\lambda\rho}u_{\nu}F_{\lambda\rho}$. 

\textit{4. Magnetic moment and energy shift}. In this section we will
derive the magnetic moment and energy shift from the formal solution
of the Wigner function (\ref{eq:1st-solution}). To this end we assume
that the chemical potential $\mu_{s}$ depends on space-time and 3-momentum.
For convenience, we will focus on the electromagnetic term in this
section and turn off the vortical term. The vortical term will be
investigated separately in the next section. 

If $\mu_{s}$ does not depend on the 3-momentum, from Eq. (\ref{eq:1st-solution}),
we obtain the fermion number current $j_{s}^{\rho}$ (the electric
current should be $Qj_{s}^{\rho}$), 
\begin{eqnarray}
j_{s}^{\rho}(x) & = & \int d^{4}p\mathscr{J}_{s}^{\rho}(x,p)=n_{s}^{(0)}u^{\rho}+\sigma_{\chi,s}^{(0)}B^{\rho},\label{eq:j0-em}
\end{eqnarray}
where $\sigma_{\chi,s}^{(0)}=sQ\frac{\mu_{s}}{4\pi^{2}}$ is the parity-odd
part of the conductivity in the static limit for chiral fermions with
chirality $s$. Since $f_{s}(x,p)$ does not depend on 3-momentum,
the momentum integral involving $\mathscr{J}_{(0)s}^{\rho}$ is non-vanishing
only along the fluid velocity $u^{\rho}$, so one can just make replacement
$p^{\rho}\rightarrow p_{0}u^{\rho}$. Here $n_{s}^{(0)}$ is the fermion
number density of a non-interacting gas of chiral fermions. For simplicity
of notation, we will work in the local rest frame of a fluid cell
with $u^{\mu}=(1,0,0,0)$, then we have $E_{p}=|\mathbf{p}|$. We
will use the velocity on-shell vector $v^{\rho}=(1,\mathbf{v})$ with
the 3-velocity $\mathbf{v}=\mathbf{p}/E_{p}$. 

In order to extract the magnetic moment and its energy shift, we can
extend our previous scenario by assuming that $\mu_{s}$ depends on
the 3-momentum since the energy shift from the magnetic moment can
be grouped into the effective chemical potential, while the magnetic
moment of a chiral fermion is proportional to its spin which is polarized
along its momentum. We can write the chemical potential into the sum
of a constant part and a ($x,\mathbf{p}$)-dependent part, $\mu_{s}^{e}(x,\mathbf{p})\approx\mu_{s}+W_{s}^{e}(x,\mathbf{p})$,
where $\mu_{s}$ do not depend on $x$ and $\mathbf{p}$, and $e=\pm1$
denote positive/negative energy corresponding to fermions $(e=+)$
and anti-fermions $(e=-)$. The quantities $W_{s}^{e}(x,\mathbf{p})$
can be further decomposed as $W_{s}^{e}(x,\mathbf{p})=W(x,\mathbf{v})+\frac{se}{2E_{p}}W_{5}(x,\mathbf{v})$,
where the introduction of the factor $1/(2E_{p})$ in the second term
is to make $W_{5}$ depends on $\mathbf{v}$ only. Note that our definitions
of $\mu_{5}$, $W$ and $W_{5}$ can be matched to those in Ref. \cite{Manuel:2013zaa}
up to constants. In this Letter, we will not use the specific form
of $\mu_{s}^{e}(x,\mathbf{p})$ except its formal dependence on $e$,
$s$, $x$ and $\mathbf{p}$. 

With $\mu_{s}^{e}(x,\mathbf{p})$, $f_{s}(x,p)$ in Eq. (\ref{eq:dist})
now become 
\begin{eqnarray}
f_{s}(x,p) & = & \frac{2}{(2\pi)^{3}}\sum_{e=\pm1}\Theta(ep_{0})f_{F}[ep_{0}-e\mu_{s}^{e}(x,e\mathbf{p})].\label{eq:fsxp}
\end{eqnarray}
Note that there is a minus sign ($e=-1$) in front of $\mathbf{p}$
in the anti-fermion sector. So the fermion number 4-current $j_{s}^{\rho}(x)$
can be obtained by integrating $\mathscr{J}_{s}^{\rho}(x,p)$ in Eq.
(\ref{eq:1st-solution}) over 4-momentum with $f_{s}$ given by Eq.
(\ref{eq:fsxp}). The fermion number density can be derived in Appendix
\ref{sec:app-a} and be put into a compact form 
\begin{eqnarray}
n_{s} & = & \int\frac{d^{3}p}{(2\pi)^{3}}\sum_{e=\pm1}ef_{F}[E_{p}-e\mu_{s}^{e}(x,\mathbf{p})]+\int\frac{d^{3}p}{(2\pi)^{3}}\frac{1}{2E_{p}^{2}}sQ(\mathbf{v}\cdot\mathbf{B})\sum_{e=\pm1}ef_{F}[E_{p}-e\mu_{s}^{e}(x,\mathbf{p})]\nonumber \\
 &  & -\int\frac{d^{3}p}{(2\pi)^{3}}\frac{1}{2E_{p}^{2}}E_{p}sQ(\mathbf{v}\cdot\mathbf{B})\frac{d}{dE_{p}}\sum_{e=\pm1}ef_{F}[E_{p}-e\mu_{s}^{e}(x,\mathbf{p})]\nonumber \\
 & \approx & \int\frac{d^{3}p}{(2\pi)^{3}}\sqrt{\gamma}\sum_{e=\pm1}ef_{F}[E_{p}^{\prime}-e\mu_{s}^{e}(x,\mathbf{p})],\label{eq:fn-density}
\end{eqnarray}
where $\sqrt{\gamma}\equiv(1+sQ\mathbf{a}_{p}\cdot\mathbf{B})$ is
the phase-space measure with the Berry curvature $\mathbf{a}_{p}=\frac{1}{2E_{p}^{2}}\mathbf{v}$,
and $E_{p}^{\prime}=E_{p}(1-sQ\mathbf{a}_{p}\cdot\mathbf{B})$ is
the effective energy. So from $E_{p}^{\prime}$ we can read out the
energy shift and magnetic moment of chiral fermions, 
\begin{eqnarray}
\Delta E_{B} & = & -\boldsymbol{\mu}_{m}\cdot\mathbf{B}\label{eq:m-energy}\\
\boldsymbol{\mu}_{m} & = & g\frac{eQ}{2|\mathbf{p}|}\mathbf{S}=\frac{sQ}{2|\mathbf{p}|}\mathbf{v}
\end{eqnarray}
where $\mathbf{S}=\frac{1}{2}es\mathbf{v}$ and $g=2$ are the spin
and g-factor of charged chiral fermions respectively. For $e=s=+$
or $e=s=-$ (positive energy and right-hand chirality or negative
energy and left-hand chirality), $\mathbf{S}=\frac{1}{2}\mathbf{v}$
(positive helicity), otherwise $\mathbf{S}=-\frac{1}{2}\mathbf{v}$
(negative helicity). However $\boldsymbol{\mu}_{m}$ only depends
on chirality and charge of positive energy fermions. For $s=Q=+$
or $s=Q=-$ (positive chirality and positive charge or negative chirality
and negative charge), $\boldsymbol{\mu}_{m}$ is parallel to $\mathbf{p}$,
otherwise it is anti-parallel to $\mathbf{p}$. We also see the emergence
of the phase-space measure $\sqrt{\gamma}$, which is not surprising
since we have already shown this in Ref. \cite{Chen:2012ca}. What
is new is the emergence of the magnetic moment and energy shift in
our Lorentz covariant kinetic approach although superficially there
is no such terms in the distribution function $f_{s}$ in Eq. (\ref{eq:fsxp}).
We then reproduce the results of Son and Yamamoto \cite{Son:2012zy,Satow:2014lva,Manuel:2013zaa}. 

\textit{5. Energy shift from spin-vorticity coupling}. We now consider
the vorticity term in the Wigner function $\mathscr{J}_{(1)s}^{\rho}$
in Eq. (\ref{eq:1st-solution}). The fermion number current from vorticity
term is given by 
\begin{eqnarray}
j_{(1)s}^{\rho}(x) & = & -\frac{s}{2}\int d^{4}p\tilde{\Omega}^{\rho\beta}p_{\beta}\frac{df_{s}}{dp_{0}}\delta(p^{2})\nonumber \\
 & = & s\frac{1}{4\pi^{2}}\left(\frac{\pi^{2}}{3}T^{2}+\mu_{s}^{2}\right)\omega^{\rho}+u^{\rho}\int\frac{d^{3}p}{(2\pi)^{3}}\sum_{e=\pm1}(-\frac{es}{2}\boldsymbol{\omega}\cdot\mathbf{v})e\frac{d}{dE_{p}}f_{F}[E_{p}-e\mu_{s}^{e}(x,\mathbf{p})],\label{eq:vorticity-shift}
\end{eqnarray}
where the first term ($\sim\omega^{\rho}$) gives the CVE coefficient,
while the second term $\sim u^{\rho}$ provides the correction to
the fermion number density from the spin-vorticity coupling energy.
Note that these two terms are orthogonal to each other since $u\cdot\omega=0$,
so there is no correction to the CVE coefficient at this level from
the second term. We note that there are a few arguments about the
temperature dependent part of the CVE coefficient such as its origin
and whether there are corrections from high order terms \cite{Jensen:2012kj,Hou:2012xg,Kalaydzhyan:2014bfa}.
The $u^{\rho}$ part of $j_{s(1)}^{\rho}$ in Eq. (\ref{eq:vorticity-shift})
gives the vorticity contribution to $n_{s}$ in Eq. (\ref{eq:fn-density}).
So the spin-vorticity energy shift can be read out from Eq. (\ref{eq:vorticity-shift}),
\begin{equation}
\Delta E_{\omega}=-\frac{1}{2}es(\boldsymbol{\omega}\cdot\mathbf{v})=-\boldsymbol{\omega}\cdot\mathbf{S},\label{eq:v-en-shift-2}
\end{equation}
where $\boldsymbol{\omega}$ is the vorticity 3-vector. From the energy
density one can also derive the correction from the vorticity term
of $\mathscr{J}_{(1)s}^{\rho}$, 
\begin{eqnarray}
\epsilon_{s(1)} & = & \int\frac{d^{3}p}{(2\pi)^{3}}E_{p}\sum_{e=\pm1}(-\frac{es}{2}\boldsymbol{\omega}\cdot\mathbf{v})\frac{d}{dE_{p}}f_{F}[E_{p}-e\mu_{s}^{e}(x,\mathbf{p})].\label{eq:en-correction-om}
\end{eqnarray}
One can also read out the same energy shift $\Delta E_{\omega}$ from
the spin-vorticity coupling as (\ref{eq:v-en-shift-2}). 

Combining Eq. (\ref{eq:m-energy}) and (\ref{eq:v-en-shift-2}), we
obtain the total effective energy with energy shifts from the magnetic
moment and the spin-vorticity coupling, 
\begin{eqnarray}
E_{p}^{\prime} & = & E_{p}-\boldsymbol{\mu}_{m}\cdot\mathbf{B}-\boldsymbol{\omega}\cdot\mathbf{S}.
\end{eqnarray}
We see the emergence of the magnetic moment and the spin-vorticity
coupling from the Wigner function solution (\ref{eq:1st-solution}). 

\textit{6. Parity-odd electric conductivity}. We can calculate the
parity-odd part of the electric conductivity or chiral magnetic conductivity,
$\sigma_{\chi}(\omega,\mathbf{k})$, from the Wigner function solution
(\ref{eq:1st-solution}) as the result of expansion in $\nabla^{\mu}=\partial_{x}^{\mu}-QF^{\mu\nu}\partial_{\nu}^{p}$,
where we assume that $\partial_{x}^{\mu}\sim(\omega,\mathbf{k})$
is of the same order as $F^{\mu\nu}$. This is valid for small $\omega$
and $|\mathbf{k}|$. For finite values of $\omega$ and $|\mathbf{k}|$,
a more rigorous way is to carry out the expansion of linear response
in space-time varying fields only and regard $\partial_{x}^{\mu}\sim(\omega,\mathbf{k})$
as $O(1)$ quantities. In this case, one should solve a different
set of equations with similar structure to Eq. (\ref{eq:wig-eq})
but with replacements $p^{\mu}\rightarrow p^{\mu}-\frac{1}{2}Qj_{1}\left(\frac{1}{2}\Delta\right)F^{\mu\nu}\partial_{\nu}^{p}$
and $\nabla^{\mu}\rightarrow\partial_{x}^{\mu}-Qj_{0}\left(\frac{1}{2}\Delta\right)F^{\mu\nu}\partial_{\nu}^{p}$,
where $\Delta\equiv\partial_{x}\cdot\partial_{p}$ and $\partial_{x}$
acts only on $F^{\mu\nu}$ but not on distributions. Here $j_{0}(y)=\sin(y)/y$
and $j_{1}(y)=(\sin y-y\cos y)/y^{2}$ are spherical Bessel functions.
The solution to the leading equations is still $\mathscr{J}_{(0)s}^{\rho}=p^{\rho}f_{s}(p)\delta(p^{2})$,
same as in Eqs. (\ref{eq:1st-solution},\ref{eq:dist}) except that
$f_{s}(p)$ does not depend on $x$ because $\mathscr{J}_{(0)s}^{\rho}$
should satisfy $\partial_{\mu}^{x}\mathscr{J}_{s(0)}^{\mu}=0$. We
refer the readers to Appendix \ref{sec:app-b} for the details of
Wigner equations and its solutions in this case. 

As shown in Appendix \ref{sec:app-b}, the parity-odd part of the
Wigner functions in $k=(\omega,\mathbf{k})$ representation which
linearly depends on fields can be written in the following form  
\begin{eqnarray}
\mathscr{J}_{(1)\mu}^{s}(k,p) & = & -i\frac{sQ}{2p\cdot k}\epsilon_{\mu\nu\rho\sigma}k^{\nu}p^{\sigma}A^{\rho}(k)j_{0}\left(\frac{1}{2}\Delta\right)(k\cdot\partial_{p})[f_{s}\delta(p^{2})]\nonumber \\
 & = & i\frac{sQ}{2p\cdot k}\epsilon_{\mu\nu\rho\sigma}k^{\nu}p^{\rho}A^{\sigma}\left\{ f_{s}\left(p+\frac{1}{2}k\right)\delta\left[\left(p+\frac{1}{2}k\right)^{2}\right]\right.\nonumber \\
 &  & \left.-f_{s}\left(p-\frac{1}{2}k\right)\delta\left[\left(p-\frac{1}{2}k\right)^{2}\right]\right\} ,\label{eq:chiral-part}
\end{eqnarray}
where we have used $\Delta=-ik\cdot\partial_{p}$, $j_{0}\left(\frac{1}{2}\Delta\right)(k\cdot\partial_{p})=2i\sin\left(\frac{1}{2}\Delta\right)$
and $\exp\left(\frac{1}{2}k\cdot\partial_{p}\right)f_{s}\delta(p^{2})=f_{s}\left(p+\frac{1}{2}k\right)\delta\left[\left(p+\frac{1}{2}k\right)^{2}\right]$.
We can obtain the 3-vector current by integration over 4-momentum
for the spatial component of $\mathscr{J}_{(1)\mu}^{s}$, i.e. $\mathbf{j}_{s}^{i}(\omega,\mathbf{k})=-\int d^{4}p\mathscr{J}_{(1),\mu=i}^{s}$.
We can then read out $\sigma_{\chi}^{s}(\omega,\mathbf{k})$ which
is just the one-loop result, Eq. (36) of Ref. \cite{Kharzeev:2009pj},
\begin{eqnarray}
\sigma_{\chi}^{s}(\omega,\mathbf{k}) & = & \frac{sQ}{16\pi^{2}}\frac{\mathbf{k}^{2}-\omega^{2}}{|\mathbf{k}|^{3}}\int d|\mathbf{p}|\left[f_{F}(|\mathbf{p}|-\mu_{s})-f_{F}(|\mathbf{p}|+\mu_{s})\right]\nonumber \\
 &  & \left[\left(2|\mathbf{p}|-\omega\right)\ln\frac{(\omega-|\mathbf{p}|)^{2}-(|\mathbf{p}|+|\mathbf{k}|)^{2}}{(\omega-|\mathbf{p}|)^{2}-(|\mathbf{p}|-|\mathbf{k}|)^{2}}\right.\nonumber \\
 &  & \left.+\left(2|\mathbf{p}|+\omega\right)\ln\frac{(\omega+|\mathbf{p}|)^{2}-(|\mathbf{p}|+|\mathbf{k}|)^{2}}{(\omega+|\mathbf{p}|)^{2}-(|\mathbf{p}|-|\mathbf{k}|)^{2}}\right].
\end{eqnarray}
If we assume that external frequency and momentum are much smaller
than internal momentum, $\omega,|\mathbf{k}|\ll|\mathbf{p}|$, we
can reproduce the HTL or HDL result \cite{Son:2012zy,Manuel:2013zaa},
\begin{eqnarray}
\sigma_{\chi}^{\mathrm{HTL/HDL}}(\omega,\mathbf{k}) & = & \sigma_{\chi}^{(0)}\left(1-\frac{\omega^{2}}{|\mathbf{k}|^{2}}\right)\left[1-\frac{\omega}{2|\mathbf{k}|}\ln\frac{\omega+|\mathbf{k}|}{\omega-|\mathbf{k}|}\right],
\end{eqnarray}
where we have used $\sigma_{\chi}^{(0)}=\sigma_{\chi,\mathrm{R}}^{(0)}+\sigma_{\chi,\mathrm{L}}^{(0)}=\frac{1}{2\pi^{2}}Q\mu_{5}$.
By including finite damping rate into fermionic propagators the non-commutativity
of the static limit $\omega\rightarrow0$ and $|\mathbf{k}|\rightarrow0$
can be removed \cite{Satow:2014lva}. 

\textit{7. Summary}. We have investigated properties of chiral fermions
in electromagnetic fields in covariant quantum kinetic theory based
on 4D Wigner functions. We have shown that the energy shifts of chiral
fermions from the magnetic moment and the spin-vorticity coupling
emerge from the Wigner function solutions derived from our previous
works. We have also calculated the parity-odd electric conductivity
in linear response theory and reproduced the results of one-loop and
HTL/HDL. All these properties, together with previous findings, show
that the 4D Wigner functions capture comprehensive aspects of physics
for chiral fermions in electromagnetic fields. 

\textit{Acknowledgments.} QW and JHG are supported in part by the
Major State Basic Research Development Program (MSBRD) in China under
Grant 2015CB856902 and 2014CB845406 respectively and by the National
Natural Science Foundation of China (NSFC) under the Grant 11125524
and 11475104 respectively. QW was supported jointly by China Scholarship
Council and the nuclear theory group of Brookhaven National Laboratory
as a senior research fellow when this work was completed. QW thanks
D. Kharzeev for many insightful discussions and good suggestions,
he also thanks S. Lin, C. Manuel, H.C. Ren, H. Yee and Y. Yin for
helpful discussions. 

\appendix

\section{Derivation of Eq. (\ref{eq:fn-density})}

\label{sec:app-a}In the case with momentum dependent chemical potential
$\mu_{s}^{e}(x,\mathbf{p})$, the zeroth order term of $j_{s}^{\rho}$
which corresponds to $\mathscr{J}_{(0)s}^{\rho}$ in Eq. (\ref{eq:1st-solution})
can be derived as 
\begin{eqnarray}
j_{(0)s}^{\rho}(x) & = & \int d^{4}pf_{s}(x,p)p^{\rho}\delta(p^{2})\nonumber \\
 & = & \int\frac{d^{3}p}{(2\pi)^{3}}\left[(u^{\rho}+\tilde{v}^{\rho})f_{F}(E_{p}-\mu_{s}^{+}(x,\mathbf{p}))+(-u^{\rho}+\tilde{v}^{\rho})f_{F}(E_{p}+\mu_{s}^{-}(x,-\mathbf{p}))\right]\nonumber \\
 & = & \int\frac{d^{3}p}{(2\pi)^{3}}v^{\rho}\left[f_{F}(E_{p}-\mu_{s}^{+}(x,\mathbf{p}))-f_{F}(E_{p}+\mu_{s}^{-}(x,\mathbf{p}))\right].\label{eq:j0-eff-1}
\end{eqnarray}
In the above equation, we have substituted $f_{s}(x,p)$ of Eq. (\ref{eq:fsxp}),
used $\delta(p^{2})=\frac{1}{2E_{p}}[\delta(p_{0}-E_{p})+\delta(p_{0}+E_{p})]$,
and carried out the integral over $p_{0}$. We have used the notation
$\tilde{v}^{\rho}\equiv(0,\mathbf{v})$ with $\mathbf{v}\equiv\mathbf{p}/E_{p}=\mathbf{p}/|\mathbf{p}|$.
From the second to the third equality, we have changed the integral
variable $\mathbf{p}\rightarrow-\mathbf{p}$ for the second term (the
negative energy sector) inside the square brackets. 

The first order term of $j_{s}^{\rho}$ corresponding to the electromagnetic
part of $\mathscr{J}_{(1)s}^{\rho}$ in Eq. (\ref{eq:1st-solution})
can be written as 
\begin{eqnarray}
j_{(1)s}^{\rho}(x) & = & sQ\tilde{F}^{\rho\lambda}\int d^{4}pf_{s}(x,p)p_{\lambda}\delta^{\prime}(p^{2})\nonumber \\
 & = & \frac{s}{2}Q\tilde{F}^{\rho\lambda}\int d^{4}pf_{s}(x,p)\frac{p_{\lambda}}{p_{0}}\frac{d}{dp_{0}}\delta(p^{2})\nonumber \\
 & = & -\frac{s}{2}Q\tilde{F}^{\rho\lambda}\int d^{4}p\frac{d}{dp_{0}}\left[f_{s}(x,p)\frac{p_{\lambda}}{p_{0}}\right]\delta(p^{2})\nonumber \\
 &  & +\frac{s}{2}Q\tilde{F}^{\rho\lambda}\int d^{4}p\frac{d}{dp_{0}}\left[f_{s}(x,p)\frac{p_{\lambda}}{p_{0}}\delta(p^{2})\right]\nonumber \\
 & = & -\frac{s}{2}Q\tilde{F}^{\rho\lambda}\int d^{4}p\left[\frac{p_{\lambda}}{p_{0}}\cdot\frac{df_{s}(x,p)}{dp_{0}}-f_{s}(x,p)\frac{\tilde{p}_{\lambda}}{p_{0}^{2}}\right]\delta(p^{2}).\label{eq:j1-eff}
\end{eqnarray}
In the first equality we have used the identities $\delta^{\prime}(p^{2})\equiv\frac{d}{dp^{2}}\delta(p^{2})=-\frac{1}{p^{2}}\delta(p^{2})$
and $\frac{d}{dp^{2}}\delta(p^{2})=\frac{d}{dp_{0}^{2}}\delta(p^{2})$.
From the third to the fourth equality, we have dropped the integral
of complete derivative since $f_{s}(x,p)\frac{p_{\lambda}}{p_{0}}\delta(p^{2})\rightarrow0$
at the boundaries $p_{0}\rightarrow\pm\infty$. In the last equality
we have used the notation $\tilde{p}^{\lambda}=(0,\mathbf{p})$. In
order to further simplify Eq. (\ref{eq:j1-eff}), we will use 
\begin{eqnarray}
\frac{df_{s}(x,p)}{dp_{0}} & = & \frac{2}{(2\pi)^{3}}\left[\Theta(p_{0})\frac{d}{dp_{0}}f_{F}(p_{0}-\mu_{s}^{+}(x,\mathbf{p}))\right.\nonumber \\
 &  & \left.-\Theta(-p_{0})\frac{d}{d(-p_{0})}f_{F}(-p_{0}+\mu_{s}^{-}(x,-\mathbf{p}))\right]\label{eq:df-dp0}
\end{eqnarray}
where we have neglected derivatives of $\Theta(\pm p_{0})$ because
they vanish when combining with $\delta(p^{2})$. Inserting Eqs. (\ref{eq:fsxp},\ref{eq:df-dp0})
into the last equality of Eq. (\ref{eq:j1-eff}), we obtain 
\begin{eqnarray}
j_{(1)s}^{\rho}(x) & = & -\frac{s}{2}Q\tilde{F}^{\rho\lambda}\int\frac{d^{3}p}{(2\pi)^{3}E_{p}}\left[(u_{\lambda}+\tilde{v}_{\lambda})\frac{df_{F}(E_{p}-\mu_{s}^{+}(x,\mathbf{p}))}{dE_{p}}+(-u_{\lambda}+\tilde{v}_{\lambda})\frac{df_{F}(E_{p}+\mu_{s}^{-}(x,-\mathbf{p}))}{dE_{p}}\right]\nonumber \\
 &  & +\frac{s}{2}Q\tilde{F}^{\rho\lambda}\int\frac{d^{3}p}{(2\pi)^{3}E_{p}^{2}}\tilde{v}_{\lambda}\left[f_{F}(E_{p}-\mu_{s}^{+}(x,\mathbf{p}))+f_{F}(E_{p}+\mu_{s}^{-}(x,-\mathbf{p}))\right]\nonumber \\
 & = & -\frac{s}{2}Q\int\frac{d^{3}p}{(2\pi)^{3}E_{p}}\tilde{F}^{\rho\lambda}v_{\lambda}\left[\frac{df_{F}(E_{p}-\mu_{s}^{+}(x,\mathbf{p}))}{dE_{p}}-\frac{df_{F}(E_{p}+\mu_{s}^{-}(x,\mathbf{p}))}{dE_{p}}\right]\nonumber \\
 &  & +\frac{s}{2}Q\int\frac{d^{3}p}{(2\pi)^{3}E_{p}^{2}}\tilde{F}^{\rho\lambda}\tilde{v}_{\lambda}\left[f_{F}(E_{p}-\mu_{s}^{+}(x,\mathbf{p}))-f_{F}(E_{p}+\mu_{s}^{-}(x,\mathbf{p}))\right],\label{eq:j1-eff-1}
\end{eqnarray}
where we have changed in the second equality the integral variable
$\mathbf{p}\rightarrow-\mathbf{p}$ for the second terms (sectors
of negative energy) inside square brackets, and we have also used
$v_{\lambda}=u_{\lambda}+\tilde{v}_{\lambda}$. 

From Eqs. (\ref{eq:j0-eff-1},\ref{eq:j1-eff-1}), we can obtain the
fermion number density by contraction of the fluid velocity $u_{\rho}$
and $j_{s}^{\rho}=j_{(0)s}^{\rho}+j_{(1)s}^{\rho}$, $n_{s}=u_{\rho}j_{s}^{\rho}(x)$,
which gives Eq. (\ref{eq:fn-density}) by using $u_{\rho}\tilde{v}^{\rho}=0$
and $u_{\rho}\tilde{F}^{\rho\lambda}v_{\lambda}=u_{\rho}\tilde{F}^{\rho\lambda}\tilde{v}_{\lambda}=\mathbf{B}\cdot\mathbf{v}$.

\section{Wigner functions in space-time varying electromagnetic fields}

\label{sec:app-b}We have discussed in section 2 and 3 that Eq. (\ref{eq:wig-eq})
holds for the constant field strength $F_{\mu\nu}$. For space-time
varying $F_{\mu\nu}$, it takes the general form (see Eqs. (5.12-5.21)
of Ref. \cite{Vasak:1987um}), 
\begin{eqnarray}
\Pi^{\mu}\mathscr{J}_{\mu}^{s} & = & 0,\nonumber \\
G^{\mu}\mathscr{J}_{\mu}^{s} & = & 0,\nonumber \\
2s(\Pi^{\mu}\mathscr{J}_{s}^{\nu}-\Pi^{\nu}\mathscr{J}_{s}^{\mu}) & = & -\epsilon^{\mu\nu\rho\sigma}G_{\rho}\mathscr{J}_{\sigma}^{s}.\label{eq:wig-eq-1}
\end{eqnarray}
The operators $\Pi^{\mu}$ and $G^{\mu}$ now replace $p^{\mu}$ and
$\nabla^{\mu}$ in Eq. (\ref{eq:wig-eq}) respectively, 
\begin{eqnarray}
\Pi^{\mu} & = & p^{\mu}-\frac{1}{2}j_{1}\left(\frac{1}{2}\Delta\right)QF^{\mu\nu}\partial_{\nu}^{p},\nonumber \\
G^{\mu} & = & \partial_{x}^{\mu}-j_{0}\left(\frac{1}{2}\Delta\right)QF^{\mu\nu}\partial_{\nu}^{p},
\end{eqnarray}
where $\Delta=\partial_{x}\cdot\partial_{p}$, and $j_{0}(x)=\sin x/x$
and $j_{1}(x)=(\sin x-x\cos x)/x^{2}$ are spherical Bessel functions.
Note that $\partial_{x}$ acts only on $F^{\mu\nu}$ but not on other
functions to its right. For constant $F_{\mu\nu}$, we recover $\Pi^{\mu}=p^{\mu}$
and $G^{\mu}=\nabla^{\mu}$ by using $j_{0}(0)=1$ and $j_{1}(0)=0$. 

The solution to Eq. (\ref{eq:wig-eq-1}) can be found by perturbation
in fields. To the first order, we can formally write 
\begin{eqnarray}
\mathscr{J}_{s}^{\mu} & = & \mathscr{J}_{(0)s}^{\mu}+\mathscr{J}_{(1)s}^{\mu}.
\end{eqnarray}
The zeroth order equations read 
\begin{eqnarray}
p^{\mu}\mathscr{J}_{(0)\mu}^{s} & = & 0,\nonumber \\
\partial_{x}^{\mu}\mathscr{J}_{(0)\mu}^{s} & = & 0,\nonumber \\
2s[p^{\mu}\mathscr{J}_{(0)s}^{\nu}-p^{\nu}\mathscr{J}_{(0)s}^{\mu}] & = & -\epsilon^{\mu\nu\rho\sigma}\partial_{\rho}^{x}\mathscr{J}_{(0)\sigma}^{s},
\end{eqnarray}
whose solution is in the same form as the first line of Eq. (\ref{eq:1st-solution})
except that $f_{s}$ does not depend on $x$ here due to $\partial_{x}^{\mu}\mathscr{J}_{(0)\mu}^{s}=0$.
The first order equations read, 
\begin{eqnarray}
p^{\mu}\mathscr{J}_{(1)\mu}^{s}-\frac{1}{2}j_{1}\left(\frac{1}{2}\Delta\right)QF^{\mu\nu}\partial_{\nu}^{p}\mathscr{J}_{(0)\mu}^{s} & = & 0,\nonumber \\
\partial_{x}^{\mu}\mathscr{J}_{(1)\mu}^{s}-j_{0}\left(\frac{1}{2}\Delta\right)QF^{\mu\nu}\partial_{\nu}^{p}\mathscr{J}_{(0)\mu}^{s} & = & 0,\nonumber \\
-\epsilon_{\mu\nu\rho\sigma}[\partial_{x}^{\rho}\mathscr{J}_{(1)s}^{\sigma}-j_{0}\left(\frac{1}{2}\Delta\right)QF^{\rho\lambda}\partial_{\lambda}^{p}\mathscr{J}_{(0)s}^{\sigma}] & = & 2s[p_{\mu}\mathscr{J}_{(1)\nu}^{s}-p_{\nu}\mathscr{J}_{(1)\mu}^{s}]\nonumber \\
 &  & +s\left[j_{1}\left(\frac{1}{2}\Delta\right)QF_{\nu\sigma}\partial_{p}^{\sigma}\mathscr{J}_{(0)\mu}^{s}\right.\nonumber \\
 &  & \left.-j_{1}\left(\frac{1}{2}\Delta\right)QF_{\mu\sigma}\partial_{p}^{\sigma}\mathscr{J}_{(0)\nu}^{s}\right].\label{eq:1st-wig-eq}
\end{eqnarray}
Contracting $\partial_{x}^{\nu}$ with the third equation and using
the second equation of Eq. (\ref{eq:1st-wig-eq}), we arrive at 
\begin{eqnarray}
p\cdot\partial_{x}\mathscr{J}_{(1)\mu}^{s}(x,p) & = & p_{\mu}j_{0}\left(\frac{1}{2}\Delta\right)QF^{\rho\sigma}\partial_{\sigma}^{p}\mathscr{J}_{(0)\rho}^{s}\nonumber \\
 &  & -\frac{1}{2}s\epsilon_{\mu\nu\rho\sigma}\partial_{x}^{\nu}\left[j_{0}\left(\frac{1}{2}\Delta\right)QF^{\rho\lambda}\partial_{\lambda}^{p}\mathscr{J}_{(0)s}^{\sigma}\right]\nonumber \\
 &  & +\frac{1}{2}Q\partial_{x}^{\nu}\left[j_{1}\left(\frac{1}{2}\Delta\right)\left(F_{\nu\sigma}\partial_{p}^{\sigma}\mathscr{J}_{(0)\mu}^{s}-F_{\mu\sigma}\partial_{p}^{\sigma}\mathscr{J}_{(0)\nu}^{s}\right)\right].
\end{eqnarray}
We can solve $\mathscr{J}_{(1)\mu}^{s}$ in momentum space by replacing
$\partial_{x}\rightarrow-ik$ and $\Delta\rightarrow-ik\cdot\partial_{p}$.
After a lengthy but straightforward calculation, we obtain 
\begin{eqnarray}
\mathscr{J}_{(1)\mu}^{s}(k,p) & = & \frac{Q}{p\cdot k}p_{\mu}[(p\cdot k)(A\cdot\partial_{p})-(p\cdot A)(k\cdot\partial_{p})]j_{0}\left(\frac{1}{2}\Delta\right)[f_{s}\delta(p^{2})]\nonumber \\
 &  & -i\frac{sQ}{2p\cdot k}\epsilon_{\mu\nu\rho\sigma}k^{\nu}p^{\sigma}A^{\rho}j_{0}\left(\frac{1}{2}\Delta\right)(k\cdot\partial_{p})[f_{s}\delta(p^{2})]\nonumber \\
 &  & +\frac{Q}{4p\cdot k}[k_{\mu}(k\cdot A)-k^{2}A_{\mu}](k\cdot\partial_{p})j_{0}\left(\frac{1}{2}\Delta\right)[f_{s}\delta(p^{2})]\nonumber \\
 &  & +i\frac{1}{2}Q[k_{\mu}(A\cdot\partial_{p})-A_{\mu}(k\cdot\partial_{p})]j_{1}\left(\frac{1}{2}\Delta\right)[f_{s}\delta(p^{2})].
\end{eqnarray}
The second term is the parity-odd part and can be put into the form
of Eq. (\ref{eq:chiral-part}). 

\bibliographystyle{apsrev}
\addcontentsline{toc}{section}{\refname}\bibliography{ref-1}

\begin{thebibliography}{26}
\expandafter\ifx\csname natexlab\endcsname\relax\def\natexlab#1{#1}\fi
\expandafter\ifx\csname bibnamefont\endcsname\relax
  \def\bibnamefont#1{#1}\fi
\expandafter\ifx\csname bibfnamefont\endcsname\relax
  \def\bibfnamefont#1{#1}\fi
\expandafter\ifx\csname citenamefont\endcsname\relax
  \def\citenamefont#1{#1}\fi
\expandafter\ifx\csname url\endcsname\relax
  \def\url#1{\texttt{#1}}\fi
\expandafter\ifx\csname urlprefix\endcsname\relax\def\urlprefix{URL }\fi
\providecommand{\bibinfo}[2]{#2}
\providecommand{\eprint}[2][]{\url{#2}}

\bibitem[{\citenamefont{Abelev et~al.}(2009)}]{Abelev:2009ac}
\bibinfo{author}{\bibfnamefont{B.}~\bibnamefont{Abelev}} \bibnamefont{et~al.}
  (\bibinfo{collaboration}{STAR Collaboration}),
  \bibinfo{journal}{Phys.Rev.Lett.} \textbf{\bibinfo{volume}{103}},
  \bibinfo{pages}{251601} (\bibinfo{year}{2009}), \eprint{0909.1739}.

\bibitem[{\citenamefont{Kharzeev et~al.}(2008)\citenamefont{Kharzeev, McLerran,
  and Warringa}}]{Kharzeev:2007jp}
\bibinfo{author}{\bibfnamefont{D.~E.} \bibnamefont{Kharzeev}},
  \bibinfo{author}{\bibfnamefont{L.~D.} \bibnamefont{McLerran}},
  \bibnamefont{and} \bibinfo{author}{\bibfnamefont{H.~J.}
  \bibnamefont{Warringa}}, \bibinfo{journal}{Nucl.Phys.}
  \textbf{\bibinfo{volume}{A803}}, \bibinfo{pages}{227} (\bibinfo{year}{2008}),
  \eprint{0711.0950}.

\bibitem[{\citenamefont{Fukushima et~al.}(2008)\citenamefont{Fukushima,
  Kharzeev, and Warringa}}]{Fukushima:2008xe}
\bibinfo{author}{\bibfnamefont{K.}~\bibnamefont{Fukushima}},
  \bibinfo{author}{\bibfnamefont{D.~E.} \bibnamefont{Kharzeev}},
  \bibnamefont{and} \bibinfo{author}{\bibfnamefont{H.~J.}
  \bibnamefont{Warringa}}, \bibinfo{journal}{Phys.Rev.}
  \textbf{\bibinfo{volume}{D78}}, \bibinfo{pages}{074033}
  (\bibinfo{year}{2008}), \eprint{0808.3382}.

\bibitem[{\citenamefont{Son and Surowka}(2009)}]{Son:2009tf}
\bibinfo{author}{\bibfnamefont{D.~T.} \bibnamefont{Son}} \bibnamefont{and}
  \bibinfo{author}{\bibfnamefont{P.}~\bibnamefont{Surowka}},
  \bibinfo{journal}{Phys.Rev.Lett.} \textbf{\bibinfo{volume}{103}},
  \bibinfo{pages}{191601} (\bibinfo{year}{2009}), \eprint{0906.5044}.

\bibitem[{\citenamefont{Kharzeev and Son}(2011)}]{Kharzeev:2010gr}
\bibinfo{author}{\bibfnamefont{D.~E.} \bibnamefont{Kharzeev}} \bibnamefont{and}
  \bibinfo{author}{\bibfnamefont{D.~T.} \bibnamefont{Son}},
  \bibinfo{journal}{Phys.Rev.Lett.} \textbf{\bibinfo{volume}{106}},
  \bibinfo{pages}{062301} (\bibinfo{year}{2011}), \eprint{1010.0038}.

\bibitem[{\citenamefont{Burnier et~al.}(2011)\citenamefont{Burnier, Kharzeev,
  Liao, and Yee}}]{Burnier:2011bf}
\bibinfo{author}{\bibfnamefont{Y.}~\bibnamefont{Burnier}},
  \bibinfo{author}{\bibfnamefont{D.~E.} \bibnamefont{Kharzeev}},
  \bibinfo{author}{\bibfnamefont{J.}~\bibnamefont{Liao}}, \bibnamefont{and}
  \bibinfo{author}{\bibfnamefont{H.-U.} \bibnamefont{Yee}},
  \bibinfo{journal}{Phys.Rev.Lett.} \textbf{\bibinfo{volume}{107}},
  \bibinfo{pages}{052303} (\bibinfo{year}{2011}), \eprint{1103.1307}.

\bibitem[{\citenamefont{Jiang et~al.}(2015)\citenamefont{Jiang, Huang, and
  Liao}}]{Jiang:2015cva}
\bibinfo{author}{\bibfnamefont{Y.}~\bibnamefont{Jiang}},
  \bibinfo{author}{\bibfnamefont{X.-G.} \bibnamefont{Huang}}, \bibnamefont{and}
  \bibinfo{author}{\bibfnamefont{J.}~\bibnamefont{Liao}}
  (\bibinfo{year}{2015}), \eprint{1504.03201}.

\bibitem[{\citenamefont{Son and Yamamoto}(2012)}]{Son:2012wh}
\bibinfo{author}{\bibfnamefont{D.~T.} \bibnamefont{Son}} \bibnamefont{and}
  \bibinfo{author}{\bibfnamefont{N.}~\bibnamefont{Yamamoto}},
  \bibinfo{journal}{Phys.Rev.Lett.} \textbf{\bibinfo{volume}{109}},
  \bibinfo{pages}{181602} (\bibinfo{year}{2012}), \eprint{1203.2697}.

\bibitem[{\citenamefont{Stephanov and Yin}(2012)}]{Stephanov:2012ki}
\bibinfo{author}{\bibfnamefont{M.}~\bibnamefont{Stephanov}} \bibnamefont{and}
  \bibinfo{author}{\bibfnamefont{Y.}~\bibnamefont{Yin}},
  \bibinfo{journal}{Phys.Rev.Lett.} \textbf{\bibinfo{volume}{109}},
  \bibinfo{pages}{162001} (\bibinfo{year}{2012}), \eprint{1207.0747}.

\bibitem[{\citenamefont{Chen et~al.}(2014{\natexlab{a}})\citenamefont{Chen,
  Pang, Pu, and Wang}}]{Chen:2013iga}
\bibinfo{author}{\bibfnamefont{J.-W.} \bibnamefont{Chen}},
  \bibinfo{author}{\bibfnamefont{J.-y.} \bibnamefont{Pang}},
  \bibinfo{author}{\bibfnamefont{S.}~\bibnamefont{Pu}}, \bibnamefont{and}
  \bibinfo{author}{\bibfnamefont{Q.}~\bibnamefont{Wang}},
  \bibinfo{journal}{Phys.Rev.} \textbf{\bibinfo{volume}{D89}},
  \bibinfo{pages}{094003} (\bibinfo{year}{2014}{\natexlab{a}}),
  \eprint{1312.2032}.

\bibitem[{\citenamefont{Gao et~al.}(2012)\citenamefont{Gao, Liang, Pu, Wang,
  and Wang}}]{Gao:2012ix}
\bibinfo{author}{\bibfnamefont{J.-H.} \bibnamefont{Gao}},
  \bibinfo{author}{\bibfnamefont{Z.-T.} \bibnamefont{Liang}},
  \bibinfo{author}{\bibfnamefont{S.}~\bibnamefont{Pu}},
  \bibinfo{author}{\bibfnamefont{Q.}~\bibnamefont{Wang}}, \bibnamefont{and}
  \bibinfo{author}{\bibfnamefont{X.-N.} \bibnamefont{Wang}},
  \bibinfo{journal}{Phys.Rev.Lett.} \textbf{\bibinfo{volume}{109}},
  \bibinfo{pages}{232301} (\bibinfo{year}{2012}), \eprint{1203.0725}.

\bibitem[{\citenamefont{Chen et~al.}(2013)\citenamefont{Chen, Pu, Wang, and
  Wang}}]{Chen:2012ca}
\bibinfo{author}{\bibfnamefont{J.-W.} \bibnamefont{Chen}},
  \bibinfo{author}{\bibfnamefont{S.}~\bibnamefont{Pu}},
  \bibinfo{author}{\bibfnamefont{Q.}~\bibnamefont{Wang}}, \bibnamefont{and}
  \bibinfo{author}{\bibfnamefont{X.-N.} \bibnamefont{Wang}},
  \bibinfo{journal}{Phys.Rev.Lett.} \textbf{\bibinfo{volume}{110}},
  \bibinfo{pages}{262301} (\bibinfo{year}{2013}), \eprint{1210.8312}.

\bibitem[{\citenamefont{Son and Yamamoto}(2013)}]{Son:2012zy}
\bibinfo{author}{\bibfnamefont{D.~T.} \bibnamefont{Son}} \bibnamefont{and}
  \bibinfo{author}{\bibfnamefont{N.}~\bibnamefont{Yamamoto}},
  \bibinfo{journal}{Phys.Rev.} \textbf{\bibinfo{volume}{D87}},
  \bibinfo{pages}{085016} (\bibinfo{year}{2013}), \eprint{1210.8158}.

\bibitem[{\citenamefont{Chen et~al.}(2014{\natexlab{b}})\citenamefont{Chen,
  Son, Stephanov, Yee, and Yin}}]{Chen:2014cla}
\bibinfo{author}{\bibfnamefont{J.-Y.} \bibnamefont{Chen}},
  \bibinfo{author}{\bibfnamefont{D.~T.} \bibnamefont{Son}},
  \bibinfo{author}{\bibfnamefont{M.~A.} \bibnamefont{Stephanov}},
  \bibinfo{author}{\bibfnamefont{H.-U.} \bibnamefont{Yee}}, \bibnamefont{and}
  \bibinfo{author}{\bibfnamefont{Y.}~\bibnamefont{Yin}},
  \bibinfo{journal}{Phys.Rev.Lett.} \textbf{\bibinfo{volume}{113}},
  \bibinfo{pages}{182302} (\bibinfo{year}{2014}{\natexlab{b}}),
  \eprint{1404.5963}.

\bibitem[{\citenamefont{Duval and Horvathy}(2014)}]{Duval:2014ppa}
\bibinfo{author}{\bibfnamefont{C.}~\bibnamefont{Duval}} \bibnamefont{and}
  \bibinfo{author}{\bibfnamefont{P.}~\bibnamefont{Horvathy}}
  (\bibinfo{year}{2014}), \eprint{1406.0718}.

\bibitem[{\citenamefont{Manuel and Torres-Rincon}(2014)}]{Manuel:2014dza}
\bibinfo{author}{\bibfnamefont{C.}~\bibnamefont{Manuel}} \bibnamefont{and}
  \bibinfo{author}{\bibfnamefont{J.~M.} \bibnamefont{Torres-Rincon}},
  \bibinfo{journal}{Phys.Rev.} \textbf{\bibinfo{volume}{D90}},
  \bibinfo{pages}{076007} (\bibinfo{year}{2014}), \eprint{1404.6409}.

\bibitem[{\citenamefont{Kharzeev and Warringa}(2009)}]{Kharzeev:2009pj}
\bibinfo{author}{\bibfnamefont{D.~E.} \bibnamefont{Kharzeev}} \bibnamefont{and}
  \bibinfo{author}{\bibfnamefont{H.~J.} \bibnamefont{Warringa}},
  \bibinfo{journal}{Phys.Rev.} \textbf{\bibinfo{volume}{D80}},
  \bibinfo{pages}{034028} (\bibinfo{year}{2009}), \eprint{0907.5007}.

\bibitem[{\citenamefont{Braaten and Pisarski}(1990)}]{Braaten:1989mz}
\bibinfo{author}{\bibfnamefont{E.}~\bibnamefont{Braaten}} \bibnamefont{and}
  \bibinfo{author}{\bibfnamefont{R.~D.} \bibnamefont{Pisarski}},
  \bibinfo{journal}{Nucl.Phys.} \textbf{\bibinfo{volume}{B337}},
  \bibinfo{pages}{569} (\bibinfo{year}{1990}).

\bibitem[{\citenamefont{Laine}(2005)}]{Laine:2005bt}
\bibinfo{author}{\bibfnamefont{M.}~\bibnamefont{Laine}},
  \bibinfo{journal}{JHEP} \textbf{\bibinfo{volume}{0510}}, \bibinfo{pages}{056}
  (\bibinfo{year}{2005}), \eprint{hep-ph/0508195}.

\bibitem[{\citenamefont{Manuel and Torres-Rincon}(2013)}]{Manuel:2013zaa}
\bibinfo{author}{\bibfnamefont{C.}~\bibnamefont{Manuel}} \bibnamefont{and}
  \bibinfo{author}{\bibfnamefont{J.~M.} \bibnamefont{Torres-Rincon}},
  \bibinfo{journal}{Phys.Rev.} \textbf{\bibinfo{volume}{D89}},
  \bibinfo{pages}{096002} (\bibinfo{year}{2013}), \eprint{1312.1158}.

\bibitem[{\citenamefont{Vasak et~al.}(1987)\citenamefont{Vasak, Gyulassy, and
  Elze}}]{Vasak:1987um}
\bibinfo{author}{\bibfnamefont{D.}~\bibnamefont{Vasak}},
  \bibinfo{author}{\bibfnamefont{M.}~\bibnamefont{Gyulassy}}, \bibnamefont{and}
  \bibinfo{author}{\bibfnamefont{H.-T.} \bibnamefont{Elze}},
  \bibinfo{journal}{Annals Phys.} \textbf{\bibinfo{volume}{173}},
  \bibinfo{pages}{462} (\bibinfo{year}{1987}).

\bibitem[{\citenamefont{Elze et~al.}(1986)\citenamefont{Elze, Gyulassy, and
  Vasak}}]{Elze:1986qd}
\bibinfo{author}{\bibfnamefont{H.-T.} \bibnamefont{Elze}},
  \bibinfo{author}{\bibfnamefont{M.}~\bibnamefont{Gyulassy}}, \bibnamefont{and}
  \bibinfo{author}{\bibfnamefont{D.}~\bibnamefont{Vasak}},
  \bibinfo{journal}{Nucl.Phys.} \textbf{\bibinfo{volume}{B276}},
  \bibinfo{pages}{706} (\bibinfo{year}{1986}).

\bibitem[{\citenamefont{Satow and Yee}(2014)}]{Satow:2014lva}
\bibinfo{author}{\bibfnamefont{D.}~\bibnamefont{Satow}} \bibnamefont{and}
  \bibinfo{author}{\bibfnamefont{H.-U.} \bibnamefont{Yee}},
  \bibinfo{journal}{Phys.Rev.} \textbf{\bibinfo{volume}{D90}},
  \bibinfo{pages}{014027} (\bibinfo{year}{2014}), \eprint{1406.1150}.

\bibitem[{\citenamefont{Jensen et~al.}(2013)\citenamefont{Jensen, Loganayagam,
  and Yarom}}]{Jensen:2012kj}
\bibinfo{author}{\bibfnamefont{K.}~\bibnamefont{Jensen}},
  \bibinfo{author}{\bibfnamefont{R.}~\bibnamefont{Loganayagam}},
  \bibnamefont{and} \bibinfo{author}{\bibfnamefont{A.}~\bibnamefont{Yarom}},
  \bibinfo{journal}{JHEP} \textbf{\bibinfo{volume}{1302}}, \bibinfo{pages}{088}
  (\bibinfo{year}{2013}), \eprint{1207.5824}.

\bibitem[{\citenamefont{Hou et~al.}(2012)\citenamefont{Hou, Liu, and
  Ren}}]{Hou:2012xg}
\bibinfo{author}{\bibfnamefont{D.-F.} \bibnamefont{Hou}},
  \bibinfo{author}{\bibfnamefont{H.}~\bibnamefont{Liu}}, \bibnamefont{and}
  \bibinfo{author}{\bibfnamefont{H.-c.} \bibnamefont{Ren}},
  \bibinfo{journal}{Phys.Rev.} \textbf{\bibinfo{volume}{D86}},
  \bibinfo{pages}{121703} (\bibinfo{year}{2012}), \eprint{1210.0969}.

\bibitem[{\citenamefont{Kalaydzhyan}(2014)}]{Kalaydzhyan:2014bfa}
\bibinfo{author}{\bibfnamefont{T.}~\bibnamefont{Kalaydzhyan}},
  \bibinfo{journal}{Phys.Rev.} \textbf{\bibinfo{volume}{D89}},
  \bibinfo{pages}{105012} (\bibinfo{year}{2014}), \eprint{1403.1256}.

\end{thebibliography}

\end{document}